\documentclass[12pt]{iopart}

\pdfoutput=1
\usepackage{graphicx}
\usepackage{epsfig}
\usepackage[latin1]{inputenc}
\usepackage{amssymb}

\begin{document}

\title[]{Near Zone Hydrodynamics of AdS/CFT Jet Wakes}

\author{Jorge Noronha$^{1}$, Miklos Gyulassy$^1$, and Giorgio Torrieri$^2$}
\address{$^1$ Department of
Physics, Columbia University, 538 West 120$^{th}$ Street, New York,
NY 10027, USA\\
$^2$ Institut f\"{u}r Theoretische Physik, Johann Wolfgang Goethe
Universit\"{a}t,\\
Max von Laue-Str. 1, 60438 Frankfurt am Main, Germany\\[0.2ex]
}

\begin{abstract}
The energy-momentum tensor of a supersonic heavy quark jet moving
through a strongly-coupled $\mathcal{N}=4$ SYM plasma is analyzed in
terms of first-order Navier-Stokes hydrodynamics. We focus on the
near zone ``head'' region close to the heavy quark
where deviations from hydrodynamic
behavior are expected to be large. For realistic quark velocities,
$v=0.99$, we find that the non-hydrodynamical head region is confined
at a narrow pancake surrounding the jet with a width
$\sim 1/T$ and transverse radius $\sim 3/T$. Outside this jet head region,
the jet induced stress is well approximated by Navier-Stokes
hydrodynamics.
\end{abstract}


%\maketitle

\section{Introduction}
The suppression of highly energetic particles observed at the
Relativistic Heavy Ion Collider (RHIC) \cite{Arsene:2004fa}
suggests  that the matter created in the first moments after the two
heavy ions have collided is a strongly-coupled deconfined plasma of
quarks and gluons \cite{Gyulassy:2004zy}. The degree of
thermalization of the soft degrees freedom seems to be consistent
with the perfect fluid hypothesis where the strongly-coupled
quark-gluon plasma (sQGP) has a shear viscosity to entropy density
ratio \cite{heinz} of the order of the string-inspired lower bound
$\eta/s=1/\left(4\pi\right)$ \cite{Policastro:2001yc} derived using
the anti-de Sitter/Conformal Field Theory (AdS/CFT) correspondence
\cite{Maldacena:1997re}. This correspondence provides a framework
to calculate nonpertubatively observables associated with
strongly-coupled $\mathcal{N}=4$ supersymmetric Yang-Mills (SYM)
theories in terms of dual weakly-coupled (super)gravity.

If the RHIC plasma is indeed strongly-coupled, one has to rely on
non-perturbative methods, such as AdS/CFT, to describe its
properties. At temperatures sufficiently above the critical
temperature, the sQGP appears to display some of the inherent
properties of $\mathcal{N}=4$ SYM plasmas such as conformal
invariance. Given the power of AdS/CFT to compute nonequilibrium
dynamical quantities, it is natural to investigate whether jet
quenching phenomena can be explained in terms of the dual
gravitational concepts of AdS/CFT.

In this paper we therefore study the SYM analogous of jet quenching, i.e., the
problem of a heavy quark moving through a strongly-coupled
$\mathcal{N}=4$ SYM plasma. We use the setup proposed in
\cite{Herzog:2006gh} where an infinitely massive heavy quark moving
at a constant speed $v$ along the comoving $X_1=X-vt$ axis in the SYM plasma corresponds to one of the
endpoints of a classical string that lives in a 5-dimensional space
defined by an AdS$_{5}$-Schwarzschild black (brane) hole geometry. In this steady-state solution, the quark
has been moving since $t \rightarrow -\infty$ and can be found at
the origin at $t=0$. According to the AdS/CFT correspondence, the disturbances in the
5-dimensional metric caused by the string can be used to obtain the
full non-perturbative result for the energy-momentum tensor of the
system plasma+heavy quark at finite temperature $T$. In Ref.\
\cite{Yarom:2007ni}, the total energy-momentum tensor of the system
was computed analytically to leading order in an expansion in powers
of inverse momentum $T/K\ll 1$ and then transformed back to coordinate
space. This tensor describes the energy and momentum disturbances
induced in the plasma in a spacetime region close to the heavy-quark
in the limit where both the t'Hooft coupling constant of the plasma
$\lambda$ and the number of colors $N_c$ are large. Note that the low-energy excitations found in the linearized analysis of Ref.\ \cite{CasalderreySolana:2004qm}, i.e, the sound and diffusion modes, only appear in the opposite limit where $T/K\gg 1$ \cite{Friess:2006fk}. In the present work, we compare the analytical result for the stress tensor derived
in \cite{Yarom:2007ni} with a first-order Navier-Stokes ansatz for
the energy-momentum tensor in order to determine the distance within
which the wake left behind by the heavy quark is thermalized (see
also \cite{Noronha:2007xe} and \cite{Chesler:2007sv}).

\section{Comparison with first-order Navier Stokes hydrodynamics}

To first order in a gradient expansion, the Navier-Stokes stress
tensor is \cite{lifshitzlandau}
\begin{equation}
\label{tmunuhydro} T^{NS}_{\mu \nu} = (\rho + p )\,U_\mu U_\nu +
p\,g_{\mu \nu} + \Pi_{\mu \nu}
\end{equation}
where $\rho$ is the local energy density, $p$ is the isotropic
pressure, and $U^\mu=(U^0,\vec{U})$ is the flow 4-vector (in our
metric $U_{\mu} U^\mu=-1$). The dissipative contribution to the
tensor above is given by the shear tensor (the bulk viscosity of
conformal plasmas is identically zero)
\begin{eqnarray}\label{shear}
\Pi^{\mu \nu} &=& -\eta\,(\partial^\mu U^\nu +
\partial^\nu U^\mu + U^\mu U_\alpha \partial^\alpha U^\nu \nonumber \\
&+&  U^\nu U^\alpha \partial_\alpha
U^\mu)+\frac{2}{3}\eta\,\Delta^{\mu\nu}\left(\partial_\alpha
U^\alpha\right),
\end{eqnarray}
where $\Delta^{\mu\nu}=g^{\mu\nu}+U^{\mu}U^{\nu}$ is the local
spatial projector and $\eta=(N_{c}-1)^2 \pi\, T^3 /8$. Note that we
kept the nonlinear terms in $U^{\mu}$ in the definition of
$\Pi_{\mu\nu}$. The flow velocity associated with the full
$T_{\mu\nu}^{Y}$ of \cite{Yarom:2007ni} can be obtained by boosting
the tensor to its local rest frame (Landau frame)
\cite{Noronha:2007xe}. The deviation from Navier-Stokes hydrodynamics can then be studied
by defining the tensor
\begin{equation}
Z_{\mu \nu} =  T_{\mu \nu}^{Y} -\Pi_{\mu\nu}  \label{zdef}
\end{equation}
and checking its components in the Landau frame. If $\left( T_{\mu
\nu}^Y \right)_L$ (components evaluated in the Landau frame come
with the L subscript) can be described by Navier-Stokes, we should
obtain
\begin{equation}
\left(Z_{11}\right)_L = \left(Z_{22}\right)_L =
\left(Z_{33}\right)_L = \frac{1}{3}
\left(Z_{00}\right)_L,\qquad\qquad\left(Z_{ij}\right)_L=0.
\end{equation}

\section{Results}

\begin{figure*}[t]
\begin{centering}
\includegraphics[width=7cm]{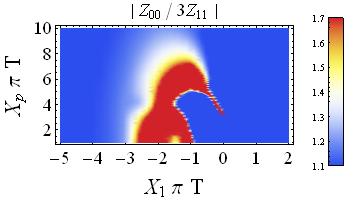}
\includegraphics[width=7cm]{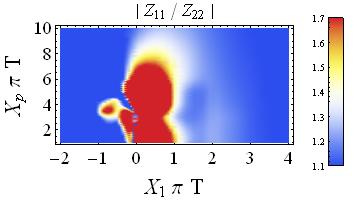}
\includegraphics[width=7cm]{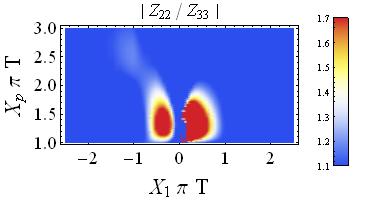}
\includegraphics[width=7cm]{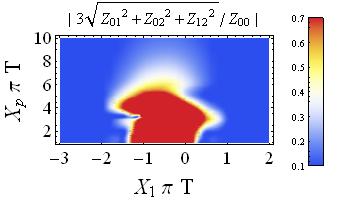}
\caption{\label{fig4} (Color online) $Z_{\mu \nu}$ is defined in Eq.
\ref{zdef}. Ratio values outside of the limits were color-coded as
the limits.}
\end{centering}
\end{figure*}

\begin{figure*}[t]
\begin{centering}
\includegraphics[width=7cm]{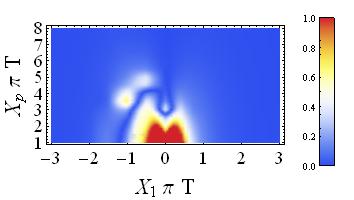} \caption{\label{fig5}
(Color online) Comparison between the full numerical solution for
$\vec{U}$ and the leading order result, i.e.,
$|\vec{U}-\vec{S}/(4P_0) |$.}
\end{centering}
\end{figure*}

One can see in Fig.\ \ref{fig4} that the magnitude and structure of
the discrepancy between Navier-Stokes and $T_{\mu\nu}^{Y}$ remains
approximately the same for all the components of $\left(Z_{\mu
\nu}\right)_{L}$. We have set $N_c=3$, $\lambda=3\pi$, and $v=0.99$.
Outside the region where $-3 < X_1 \pi T < 3$ (longitudinal
direction) and $0<X_p \pi T<10$ (transverse direction), the system
can be described by the first-order Navier-Stokes energy-momentum
tensor to an accuracy of roughly $90 \%$. Inside that region,
however, the discrepancy quickly increases and eventually diverges. In Fig.\ \ref{fig5},
one can see that in the region where Navier-Stokes provides a good
description of the system the numerically calculated velocity flow
$\vec{U}$ matches the leading order result in $1/N_c$,
$\vec{U}=\vec{S}/\left(4P_0\right)$, where $S_{i}=-T_{0i}^{Y}$ and
$P_0=\eta\,\pi T$. Therefore, in this region the system is more
accurately described by a linearized version of the first-order
Navier-Stokes equations.

%% %%%%%%%%%%%%%%%%%%%%%%%%%%
%\begin{figure*}[t]
%\epsfig{width=8cm,clip=,figure=fig2aZ003Z11v099lambda.eps}
%\epsfig{width=8cm,clip=,figure=fig2bZ11Z22v099lambda.eps}
%\epsfig{width=8cm,clip=,figure=fig2cZ22Z33v099lambda.eps}
%\epsfig{width=8cm,clip=,figure=fig3b3Z01Z02Z12Z11v099lambda.eps}
%\caption{\label{fig4} (Color online) $Z_{\mu \nu}$ is defined in Eq.
%\ref{zdef}. Ratio values outside of the limits were color-coded as
%the limits.}
%\end{figure*}
%% %%%%%%%%%%%%%%%%%%%%%

%% %%%%%%%%%%%%%%%%%%%%%%%%%%
%% %%%%%%%%%%%%%%%%%%%%%

%%%%%%%%%%%%%%%%%%%%%%
%%%%%%%%%%%%%%%%%%%%%

%%%%%%%%%%%%%%%%%%%%%%%%%%%%%%%%%%%%%%%%%%%%%%%%%%%%%%%%%%%%%%%%%%%%%%%%%%%%%%

\section{Conclusions}

In this paper we have compared the analytically obtained
energy-momentum tensor of an infinitely massive quark moving through
a strongly coupled $\mathcal{N}=4$ SYM plasma with a non-ideal
tensor given by the first-order Navier-Stokes ansatz. We have found
that the energy deposited by the quark is thermalized in the region defined by $|X_1| \pi
T>3$ and $X_p \pi T>1$, which excludes the nonequilibrium jet head. In fact, the system can be described by
linearized first-order Navier Stokes hydrodynamics down to distances
of roughly $3/(\pi T)$. Thus, assuming that a strongly coupled
$\mathcal{N}=4$ SYM plasma can capture the main features of the
sQGP, we should expect that the energy deposited by jets induces
almost perfect fluid hydrodynamic response of the sQGP outside the
head region.

%Please include your grant information !!!!!!%
\ack We thank A. Yarom, S. Gubser, and S. Pufu for useful
discussions. G.T. thanks the Alexander Von Humboldt foundation for
the support provided for this research. J.N. and M.G. acknowledge
partial support from DOE under Grant No. DE-FG02-93ER40764 and the
ITP and FIAS of the J.W. Goethe University, Frankfurt.

\end{document}